\documentclass[aps,pre,preprint,superscriptaddress,showpacs,showkeys]{revtex4-1}

\usepackage{graphicx}
\usepackage{amsmath}
\usepackage{hyperref}

\begin{document}

%\preprint{}

%\title{Existence and stability of elementary vortex knots in the Gross-Pitaevskii model}
\title{Vortex knots in a Bose-Einstein condensate}



\author{Davide Proment}
\email{davideproment@gmail.com}
\homepage{www.to.infn.it/~proment}
\affiliation{Dipartimento di Fisica Generale, Universit\`{a} degli Studi di Torino, Via Pietro Giuria 1, 10125 Torino, Italy, EU}
\affiliation{INFN, Sezione di Torino, Via Pietro Giuria 1, 10125 Torino, Italy, EU}

\author{Miguel Onorato}
\affiliation{Dipartimento di Fisica Generale, Universit\`{a} degli Studi di Torino, Via Pietro Giuria 1, 10125 Torino, Italy, EU}
\affiliation{INFN, Sezione di Torino, Via Pietro Giuria 1, 10125 Torino, Italy, EU}

\author{Carlo F. Barenghi}
\affiliation{School of Mathematics and Statistics, Newcastle University, Newcastle upon Tyne, NE1 7RU, UK, EU}

%\author{}
%\thanks{}
%\affiliation{}


\date{\today}

\begin{abstract}
We present a method for numerically building a vortex knot state in the
superfluid wave-function of a Bose-Einstein condensate. 
We integrate in time the governing Gross-Pitaevskii equation to
determine evolution and stability
of the two (topologically) simplest vortex knots which can be wrapped
over a torus.  We find that the velocity of a vortex knot
depends on the ratio of poloidal and toroidal radius:
for smaller ratio, the knot travels faster. Finally, we show how unstable 
vortex knots break up into vortex rings.
\end{abstract}



%\maketitle must follow title, authors, abstract, \pmics (fluid flow), 47.32.C-acs, and \keywords
\maketitle

\section{Introduction}

In 1867, following the works
of Helmholtz on vortices and of Riemann on 
abelian functions, Lord Kelvin modelled atoms
as knotted vortex tubes in the ether \cite{kelvin1869}, 
effectively giving birth to
knot theory \cite{adams1994knot}. This discipline has
fascinated mathematicians and physicists since.
More recently, knots have been the studied in different 
branches of physics, ranging from excitable media \cite{PhysRevE.68.016218}, 
and classical field theory \cite{Faddeev1997}, to optics \cite{Leach2004, 
Dennis2010} and liquid-crystal colloids \cite{Tkalec01072011}.

Knots in superfluids are identified with closed vortex lines, 
regions of fluid around which the circulation 
assumes non-zero (quantized) value.
Vortex rings have been studied experimentally
in superfluid liquid helium \cite{donnelly1991, PhysRevLett.100.245301}
and in Bose-Einstein condensates \cite{PhysRevLett.86.2926}. Numerical
simulations have revealed that superfluid turbulence 
contains linked vortex lines \cite{poole:10.1023}, 
but, to the best of our knowledge, individual vortices with 
non-trivial topology have never been observed directly.
To shed light onto this problem, energy, motion and stability of vortex knots 
have been examined theoretically and numerically using the classical
theory of thin-cored vortex filaments. In this approach,
the governing incompressible
Euler dynamics is expressed by the Biot-Savart law or by its
local induction approximation (LIA) \cite{ricca:1999, maggioni:2010}. 
Under certain conditions, it is found that some vortex knots
are structurally stable, that is to say they 
travel without breaking up for distance larger than their own diameters.

In superfluid helium, the validity of the classical theory of thin-core 
vortex filaments is based on the large separation of scales between the
vortex core radius $a_0$ (approximately $10^{-8}~\rm cm$ in $^4$He and
$10^{-6}~\rm cm$ in $^3$He-B) and the typical distance $\ell$ between
vortices. In turbulence experiments, $\ell \approx 10^{-3}$ to 
$10^{-4}~\rm cm$; the last value is also the typical diameter of 
experimental vortex
rings \cite{PhysRevLett.100.245301}.
The situation is very
different in atomic Bose-Einstein condensates, where $\ell$
is only few times larger than $a_0$. 
In this context, the Gross-Pitaevskii Equation (GPE) is clearly a more
realistic model \cite{pitaevskii2003bose}, particularly at very low
temperatures, as thermal effects can be neglected.

The advantage of the GPE is that it does
not need the cut-off parameter required by the classical vortex filament
theory to de-singularise the Biot-Savart integral 
\cite{saffman1991vortex}.
The second advantage is that
the GPE naturally describes vortex reconnections
\cite{PhysRevLett.71.1375}, which
must be implemented algorithmically in the
Biot-Savart model.
Any prediction about the stability and the break-up of a vortex structure
which is not orders of magnitude bigger than $a_0$ 
is therefore more reliable if obtained using the GPE.
The third advantage of searching for vortex knots
in a Bose-Einstein condensate is that direct images of individual
vortex structures are possible without the use of tracer particles
which will certainly disturb these structures. The disadvantage is that atomic
condensates are small, thus the motion of these structures will be
affected by the boundaries and by the non-uniformity density of the
background condensate. Before investigating these effects, however,
it is essential to establish
whether vortex knot solutions of the governing GPE exist, and, if 
they do, if they are sufficiently stable.  This is the limited
aim which we set in this work. We stress that we
do not aim to propose a mechanism to experimentally
create vortex knots in condensates, but only to study 
the possible existence and stability of these solutions of the GPE.
We shall see that even setting up numerically
a topologically non-trivial structure in the wave-function
is not a minor task; indeed, to the best of our knowledge, 
this is the first time it has been done. 

The manuscript is organized as the following. 
Section \ref{sec:IC} explains how to create an elementary vortex knot in
the initial conditions of the condensate wave-function. 
Section \ref{sec:DIN} deals with the analysis of the dynamical 
properties of vortex knots. Section \ref{sec:BC} describes the break-up of
vortex knots.
Finally, the conclusions are in Section \ref{sec:CONCL}.
%\clearpage

\section{Vortex knot Initial conditions \label{sec:IC}}
We consider the GPE 
written  in the following dimensionless form
\begin{equation}
2 i \, \partial_t \psi - \nabla^2 \psi + |\psi|^2 
\psi = 0,
\end{equation}
where no external confining potential is present.
The characteristic length scale of perturbations of the uniform
condensate, called {\it healing length}, is defined as
\begin{equation}
\xi = \frac{1}{\sqrt{\langle\rho\rangle}}, \,\,\, \mbox{where} \,\,\, \langle\rho
\rangle=\frac{1}{V} \int_V |\psi|^2 dV
\label{eq:healing}
\end{equation}
is the mean density of the condensate.
Besides the energy, the GPE conserves the total number of particles, therefore
$\xi$ is a conserved quantity too.
Without loss of generality, we choose to deal with a system that has an 
unperturbed density (the density field at infinity) equal to unity, and 
assume that perturbations are localized in a small region of the sample.
In this hypothesis, $ \xi \simeq 1 $ in our units.

\begin{figure}
\includegraphics[scale=1.3]{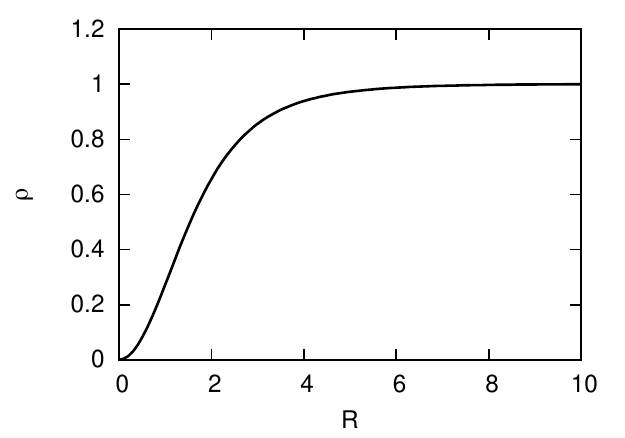}
\caption{(Colours online). The density field $\rho$ around an axisymmetric 
two-dimensional vortex. 
Radial distances $R$ are in units of the healing length $ \xi $. 
\label{fig:2DvortexDensity}}
\end{figure}

We now explain how to numerically build a vortex knot. 
First we construct a vortex. Consider the two dimensional plane $ sOz $. 
A stable vortex is a hole (zero value) in the density field, around which 
the phase of the wave-function changes by $ \pm 2\pi $.
A sufficiently accurate description of a two-dimensional vortex
centered in the origin of the $ sOz $ plane is given by the wave-function 
$ \Psi_{2D}(s, z)=\sqrt{\rho(R)} \, e^{i\theta(s, z)} $, where $ R=\sqrt{s^2+z^2} $,
\begin{equation}
\begin{split}
& \rho(R) = \frac{ R^2 \left( a_1 + a_2 R^2 \right) }
{1 + b_1 R^2 + b_2 R^4}, \\
& \theta(s, z)= \arctan \left(\frac{z}{s}\right),
\end{split}
\label{eq:vortex2D}
\end{equation}
and the coefficients $ a_1=11/32 $, $ a_2=11/384 $, $ b_1=1/3 $, 
and $ b_2=11/384 $ 
arise from a second order Pad{\'e} approximation \cite{berloff:2004}.
Fig. \ref{fig:2DvortexDensity} shows how the density field behaves
around the axisymmetric vortex centre.
It is clear from the plot that the vortex core is of the order of 
the healing length, and
the bulk value of the density $\rho=1$ is recovered at larger
distances.

We now come back to vortex knots in a three-dimensional system.
We define a knot as a closed curve  over a torus,
characterized by the toroidal radius $ R_0 $  and the poloidal radius $ R_1$.
More precisely, a closed curve $ \mathcal{T}_{n, m} $ on the torus 
is determined by 
counting  the number of toroidal wraps, $ n $, and the number of poloidal
wraps, $ m $.
For example, the curves $ \mathcal{T}_{1, 1}$ and $ \mathcal{T}_{2, 2} $ 
describe respectively the unknot (the simple vortex ring) and
two unlinked rings. The first topologically non-trivial curve is the trefoil,
$ \mathcal{T}_{2, 3}$.
In this work we shall focus on the two simplest knots, 
the trefoil $ \mathcal{T}_{2, 3} $ and its dual $ \mathcal{T}_{3, 2} $.

\subsection{The $ \mathcal{T}_{2, 3} $ knot (trefoil)}
\begin{figure}
\includegraphics[scale=0.1]{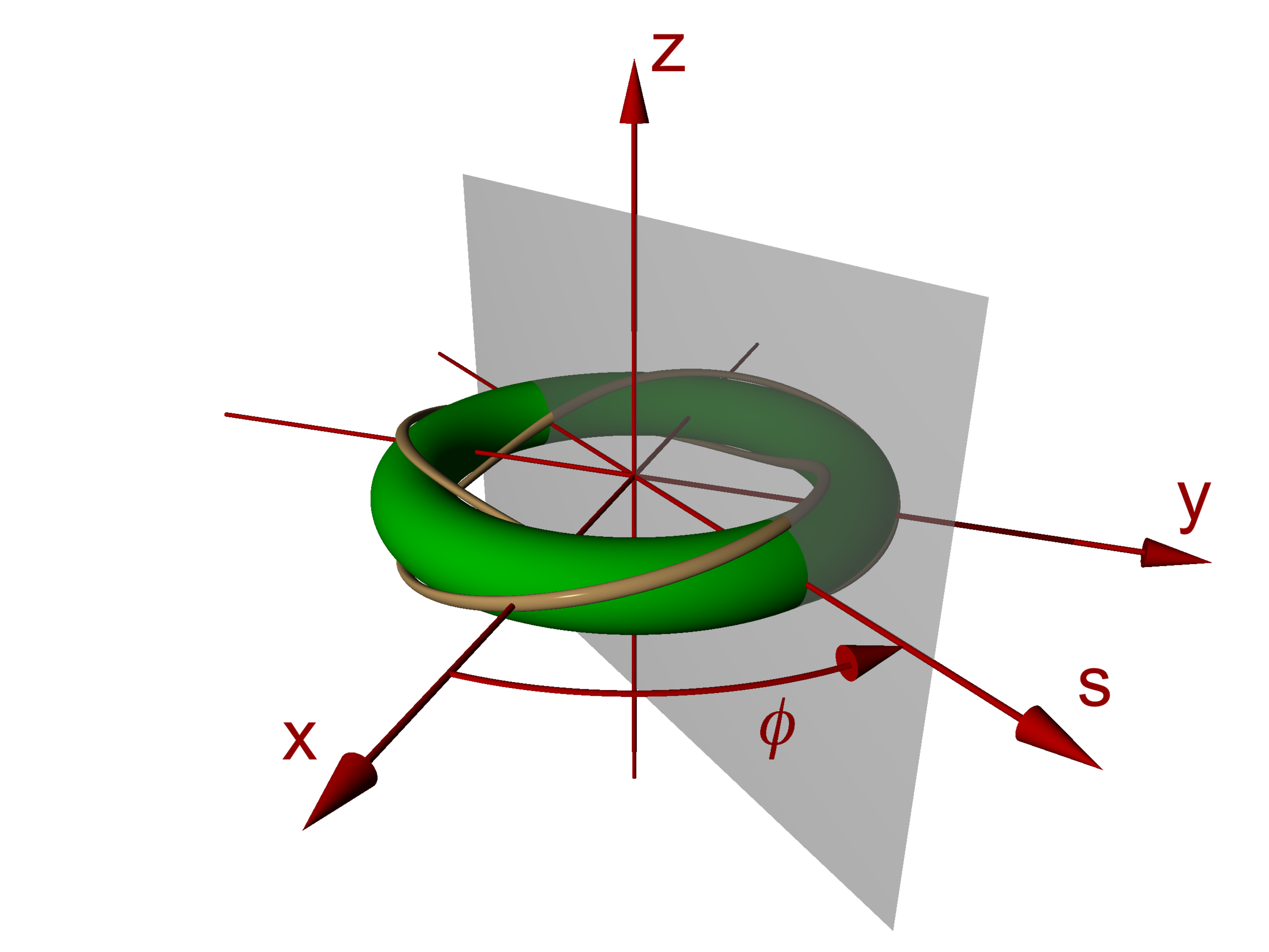}
\caption{(Colours online).
Construction of the trefoil knot $ \mathcal{T}_{2, 3} $.}
\label{fig:sampleT23}
\end{figure}

\begin{figure}
\includegraphics[scale=0.09]{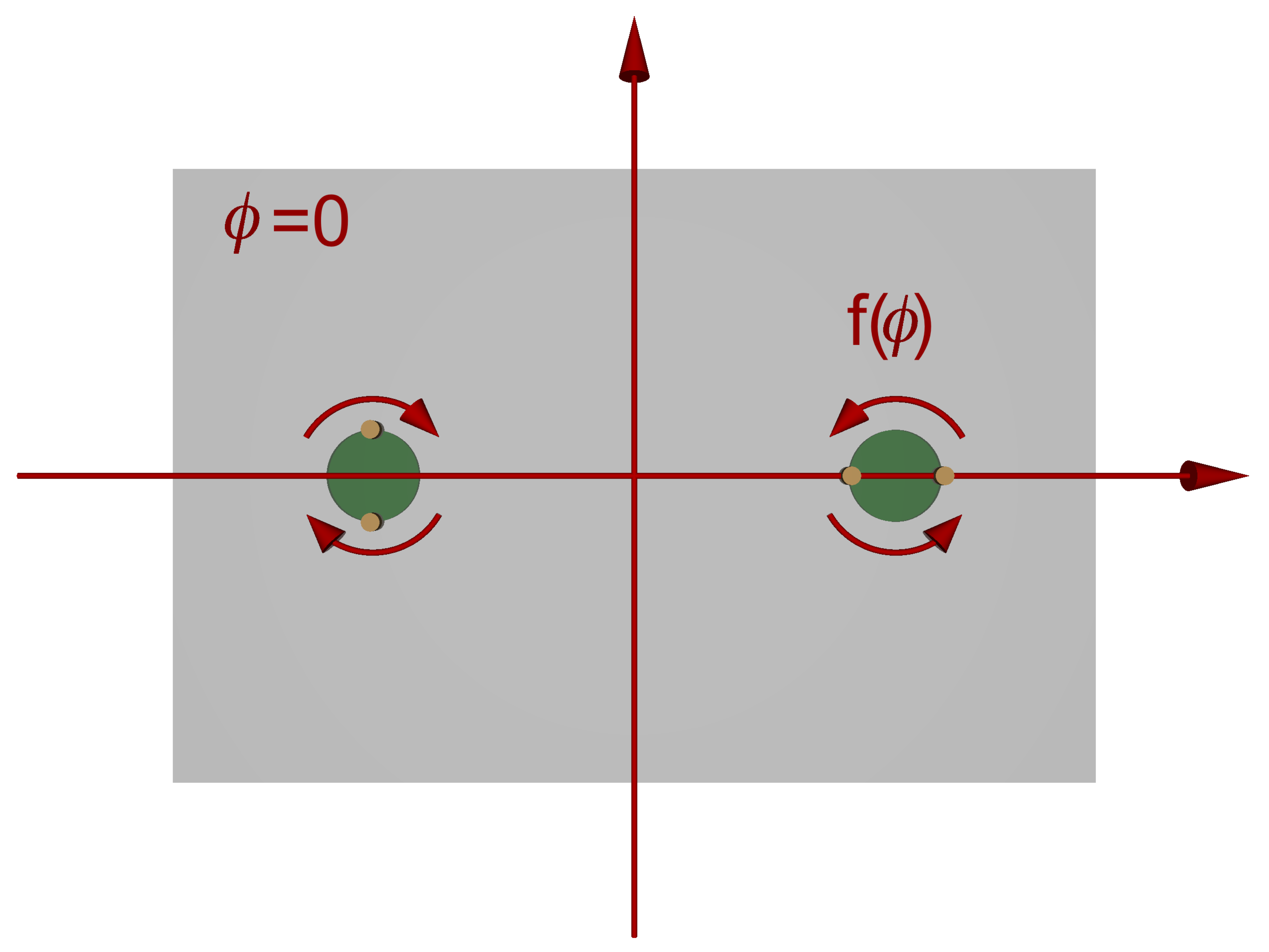}
\caption{
(Colours online).
Positions of the four vortices on $\phi=0$ used
to construct the wave-function of 
the trefoil knot $ \mathcal{T}_{2, 3} $.
\label{fig:vortexPositionsT23}}
\end{figure}

The vortex line of a $ \mathcal{T}_{2, 3} $ knot  lays on the torus  as shown
in Fig. \ref{fig:sampleT23}.
Any plane $ sOz $ passing through the $z$ axis intercepts
the curve $ \mathcal{T}_{2, 3} $ at four different points, 
which correspond to four two-dimensional point vortices on
the plane $ sOz $.
The positions of these two-dimensional vortices vary with respect to the 
choice of the plane $ sOz $; in other words, these positions
are functions of the 
angle variable $ \phi $ introduced in Fig. \ref{fig:sampleT23}.
For example, the vortex positions of the wave-function for
the angle $ \phi=0 $ are shown in Fig. \ref{fig:vortexPositionsT23}.
By construction, these point vortices are located on the circumference
defined by the intersection of the plane with the torus, and rotate on it 
following a particular function $ f(\phi) $. 
To assure continuity of the vortex line and to describe the trefoil 
knot, the function $f(\phi)$ must have the form 
$ f(\phi)=3\phi/2 $, with $ \phi \in [0, \pi) $.

We are now ready to write the three-dimensional wave-function which
describes the trefoil knot $ \mathcal{T}_{2, 3} $. 
In the approximation that the healing length $ \xi $ is much smaller
than the inter-vortex distance, the two-dimensional wave-function 
in the plane $ sOz $ is given by the superposition (multiplication)
of the wave-function $ \Psi_{2D} $ of each two-dimensional vortex 
centred in the correct position, where the opposite circulation is 
obtained by applying the complex conjugation operator $ (...)^\ast $.
Thus, the three-dimensional wave-function results in
\begin{widetext}
\begin{equation}
\begin{split}
\psi_{2, 3}(x, y, z) = \, &
\Psi_{2D}\left\{ s(x, y) - R_0 - R_1 \cos \left[\alpha(x, y)\right], z - R_1 \sin \left[\alpha(x, y)\right] \right\}
\\ \times \, &
\Psi_{2D}\left\{ s(x, y) - R_0 - R_1 \cos \left[\alpha(x, y)+\pi\right], z - R_1 \sin \left[\alpha(x, y)+\pi\right] \right\} 
\\ \times \, &
\Psi_{2D}^\ast \left\{ s(x, y) + R_0 - R_1 \cos \left[\alpha(x, y)+\pi/2\right], z + R_1 \sin \left[\alpha(x, y)+\pi/2\right] \right\}
\\ \times \, &
\Psi_{2D}^\ast\left\{ s(x, y) + R_0 - R_1 \cos \left[\alpha(x, y)+3\pi/2\right], z + R_1 \sin \left[\alpha(x, y)+3\pi/2\right] \right\},
\end{split}
\label{eq:T23}
\end{equation}
\end{widetext}
with $ s(x, y)=\mbox{sgn}(x) \sqrt{x^2+y^2} $, where $ \mbox{sgn}(...) $ 
is the sign function, and $ \alpha(x, y)=3/2 \arctan{(y/x)} $.

\subsection{The $ \mathcal{T}_{3, 2} $ knot}
\begin{figure}
\includegraphics[scale=0.1]{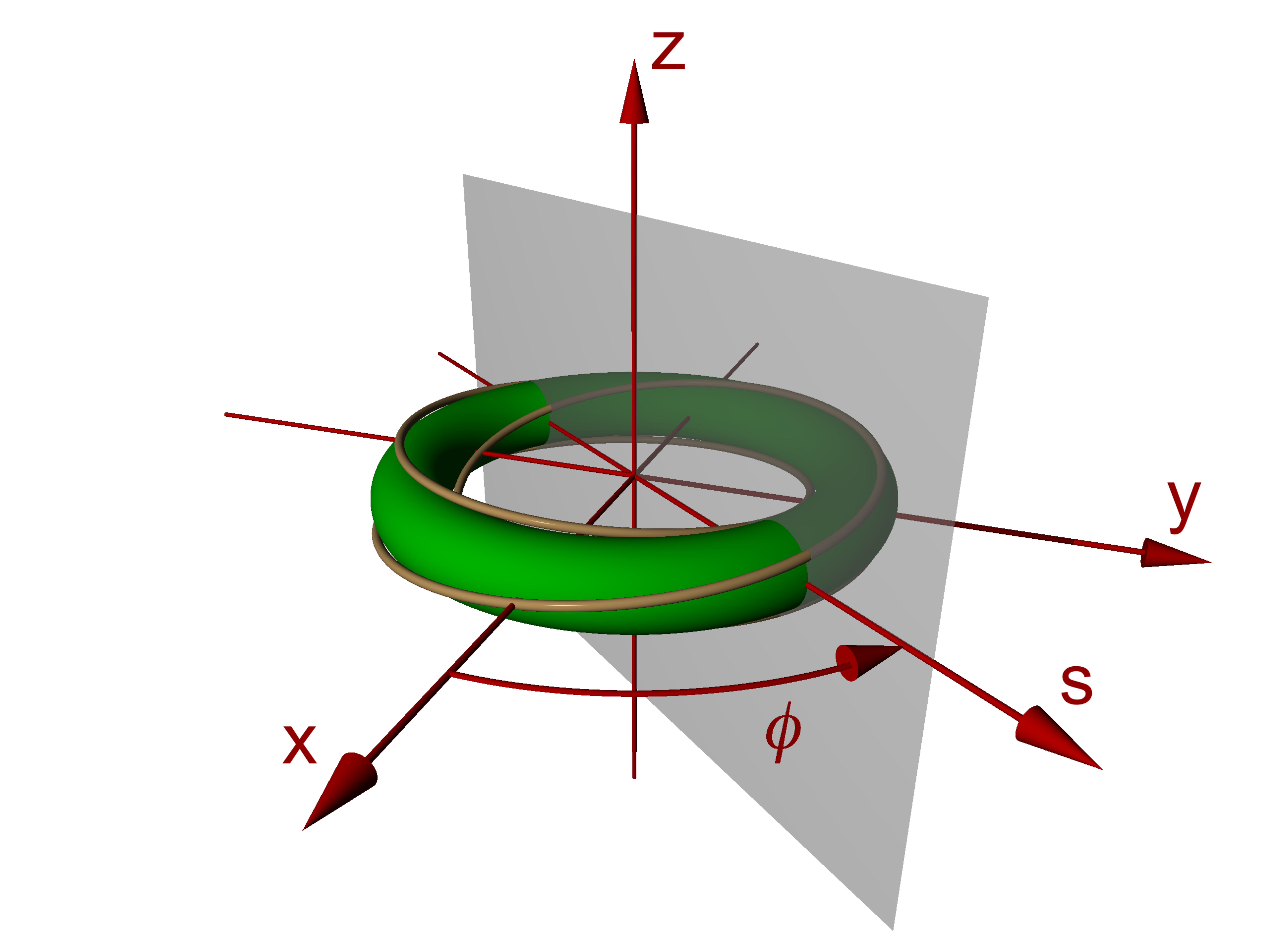}
\caption{
(Colours online).
Construction of the trefoil knot $ \mathcal{T}_{3, 2} $.
}
\label{fig:sampleT32}
\end{figure}

\begin{figure}
\includegraphics[scale=0.09]{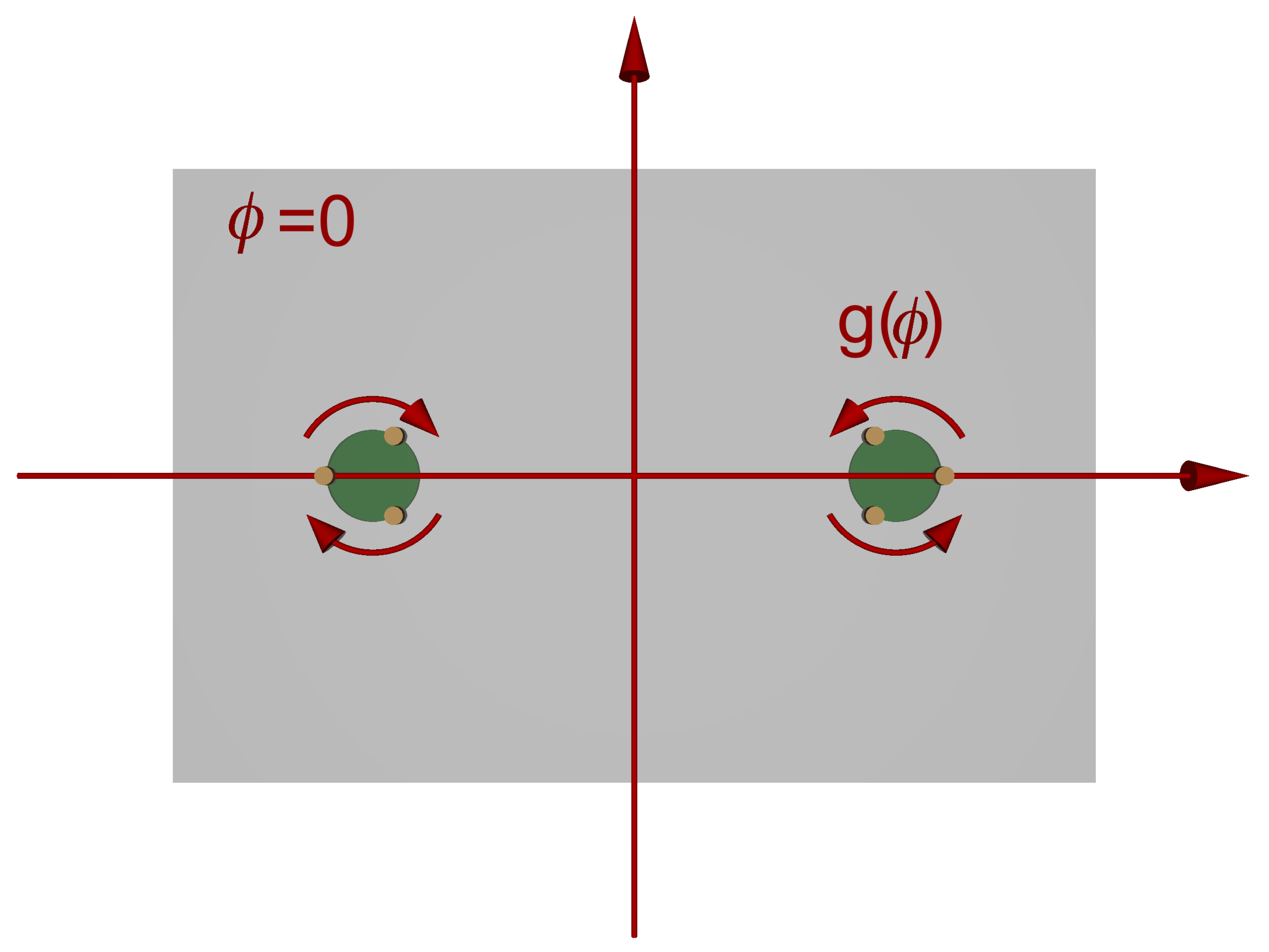}
\caption{(Colours online).
Positions of the six vortices on $\phi=0$ used to construct
the wave-function of 
the trefoil knot $ \mathcal{T}_{3, 2}$.
}
\label{fig:vortexPositionsT32}
\end{figure}

The technique used to define the wave-function of the trefoil knot can
be extended to any other knot built on a torus.
The $ \mathcal{T}_{3, 2} $ knot can be represented on the torus 
as shown in Fig. \ref{fig:sampleT32}.
In this case the generic plane $ sOz $ intersects the knot in six points,
where the centers are function $ g(\phi) $ of the angle $ \phi $ and 
rotate around the circumference defined by the plane and the torus 
intersection.
An example of the configuration for $ \phi=0 $ is shown in Fig.
\ref{fig:vortexPositionsT32}.
The function $g(\theta)$ is $ g(\theta)=2 \theta /3 $,
with $ \phi \in [0, \pi) $.

Again, using the two-dimensional vortex description 
$ \Psi_{2D} $, in the limit of inter-vortex distance much greater
than the healing length $ \xi $, the three-dimensional wave-function
of a $ \mathcal{T}_{2, 3} $ knot is
\begin{widetext}
\begin{equation}
\begin{split}
\psi_{3, 2}(x, y, z) = \, &
\Psi_{2D}\left\{ s(x, y) - R_0 - R_1 \cos \left[\alpha(x, y)\right], z - R_1 \sin \left[\alpha(x, y)\right] \right\}
\\ \times \, &
\Psi_{2D}\left\{ s(x, y) - R_0 - R_1 \cos \left[\alpha(x, y)+2\pi/3\right], z - R_1 \sin \left[\alpha(x, y)+2\pi/3\right] \right\} 
\\ \times \, &
\Psi_{2D}\left\{ s(x, y) - R_0 - R_1 \cos \left[\alpha(x, y)+4\pi/3\right], z - R_1 \sin \left[\alpha(x, y)+4\pi/3\right] \right\} 
\\ \times \, &
\Psi_{2D}^\ast\left\{ s(x, y) + R_0 + R_1 \cos \left[\alpha(x, y)\right], z + R_1 \sin \left[\alpha(x, y)\right] \right\}
\\ \times \, &
\Psi_{2D}^\ast \left\{ s(x, y) + R_0 + R_1 \cos \left[\alpha(x, y)+2\pi/3\right], z + R_1 \sin \left[\alpha(x, y)+2\pi/3\right] \right\}
\\ \times \, &
\Psi_{2D}^\ast \left\{ s(x, y) + R_0 + R_1 \cos \left[\alpha(x, y)+4\pi/3\right], z + R_1 \sin \left[\alpha(x, y)+4\pi/3\right] \right\},
\end{split}
\label{eq:T32}
\end{equation}
\end{widetext}
with $ s(x, y)=\mbox{sgn}(x) \sqrt{x^2+y^2} $ and 
$ \alpha(x, y)=2/3 \arctan{(y/x)} $.

%\clearpage

\section{Vortex knot dynamics \label{sec:DIN}}
%The parameters identifying the vortex knot are the toroidal 
%$ R_0 $ radius and the poloidal $ R_1 $ radius.
%The length scale in the system is the {\it healing length} $ \xi $,
%therefore we give a measure of these radii in terms of $ \xi $.
To study the dynamic and stability of the knots 
$ \mathcal{T}_{2, 3} $ and $ \mathcal{T}_{3, 2} $ with different 
geometries we need to find a compromise between the accessible
numerical resolution and the box size: we need to resolve small 
scales near the vortex cores and, at the same time, minimize the 
finite size (boundary) effects.
We recall that the parameters which identify our vortex knots,
the toroidal and poloidal radii $R_0$ and $R_1$, are expressed
in units of the healing length $\xi$.

We chose to uniformly discretize physical space using a
Cartesian grid with steps
$ \Delta x=\Delta y=\Delta z=0.5\, \xi $ spanning over the 
{\it knot ratios} $ R_1/R_0 = 1/10, 1/5, 2/5, 3/5 $. 
We expect vortex knots to behave similarly to vortex rings, that is
to say we expect that they travel along the direction of the torus
axis of symmetry (the $ z $-axis).
Taking our computational constraints into account, we use
$ 192 \times 192 \times 512$ grid points ($ L_x=L_y=96\, \xi $ 
and $ L_z = 256 \, \xi $) and the toroidal radius $ R_0 = 20 \, \xi $.
This choice allows us to have a minimum value of $ R_1=2 \, \xi $
(when $ R_1/R_0=1/10 $), acceptable to observe the small 
inter-vortex interactions,  and a maximum knot
size of $ 2(R_0+R_1)=64\,\xi $ (when $ R_1/R_0=3/5 $) 
which gives tolerable boundary effects.
Table \ref{tab:param} summarizes the simulation
parameters.
\begin{table}[b]
\begin{ruledtabular}
\begin{tabular}{ccccc}
case  & knot ratio & max size & min size & stable \\
$ \mathcal{T}_{2, 3} $ | $ \mathcal{T}_{3, 2} $ & $ R_1/R_0 $ & $ 2(R_0+R_1) $ & $ 2 \, R_1 $ & Yes/No \\ \hline
(a) | (e) & $ 1/10 $ & $ 44 \, \xi $ & $ 4 \, \xi $ & Y | Y \\
(b) | (f) & $ 1/5 $ & $ 48 \, \xi $ & $ 8 \, \xi $ & Y | Y  \\
(c) | (g) & $ 2/5 $ & $ 56 \, \xi $ & $ 16 \, \xi $ & N | N  \\
(d) | (h) & $ 3/5 $ & $ 64 \, \xi $ & $ 24 \, \xi $ & N | N   \\
\end{tabular}
\end{ruledtabular}
\caption{Vortex knot parameters of $ \mathcal{T}_{2, 3} $ and 
$ \mathcal{T}_{3, 2} $ used in the simulations.
\label{tab:param}}
\end{table}

In order to let the knot travel for  the maximum distance in the $ z $ 
direction, at the start of the calculation ($t=0$)
the  vortex knot is centered at the 
point $ \left[L_x/2, L_y/2, 2(R0+R1)\right] $.
With this choice, the knot can propagate for a distance of $ 3 $
to $ 53/11\simeq 4.8 $ times its maximum size, before hitting 
the opposite side of the computational domain corresponding to $ z=L_z $.
The GPE is integrated in time using a split-step 
method with anti-periodic (reflective) boundary conditions. 
The integration time step is $ \Delta t = 0.02 $ smaller that 
the fastest linear period $ T_c \simeq 0.032 $.
This value allows us to conserves the initial energy and mass
up to 3\% and 1\% respectively in all the simulations.
Details on the numerical algorithm can be found
in Ref.~\cite{Proment2011}.

\begin{figure}
\includegraphics[scale=0.2]{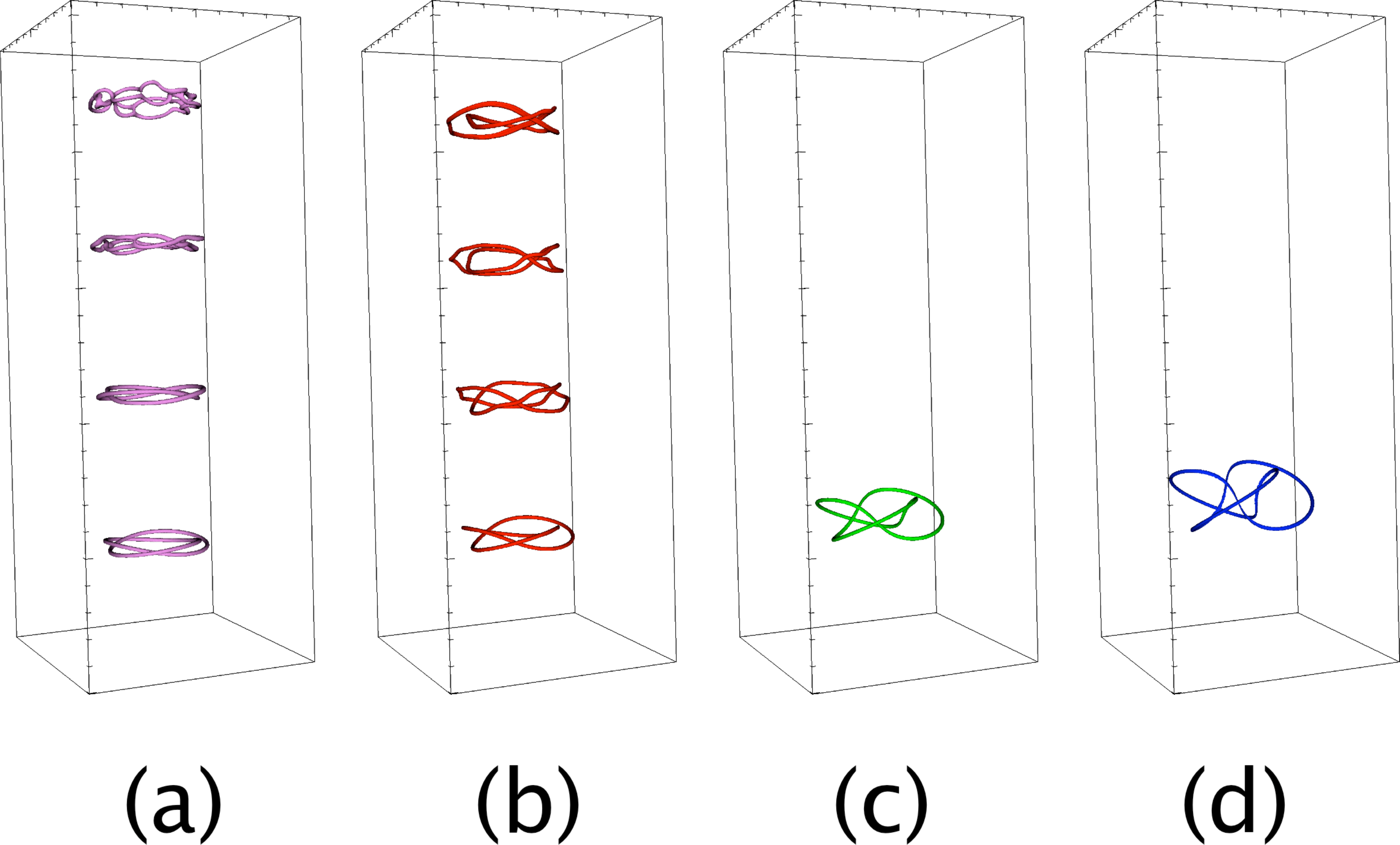}
\caption{(Colours online).
Iso-surfaces of the density field 
at the threshold level $ \rho_{th}=0.2 $ for $ \mathcal{T}_{2, 3} $ 
knots of various knot ratios $ R_1/R_0 $ (see Table 
\ref{tab:param}).
Snapshots at times $ t=0, 400,
800, 1200 $. Unstable knots are not shown.
\label{fig:evolutionT23}}
\end{figure}

\begin{figure}
\includegraphics[scale=0.2]{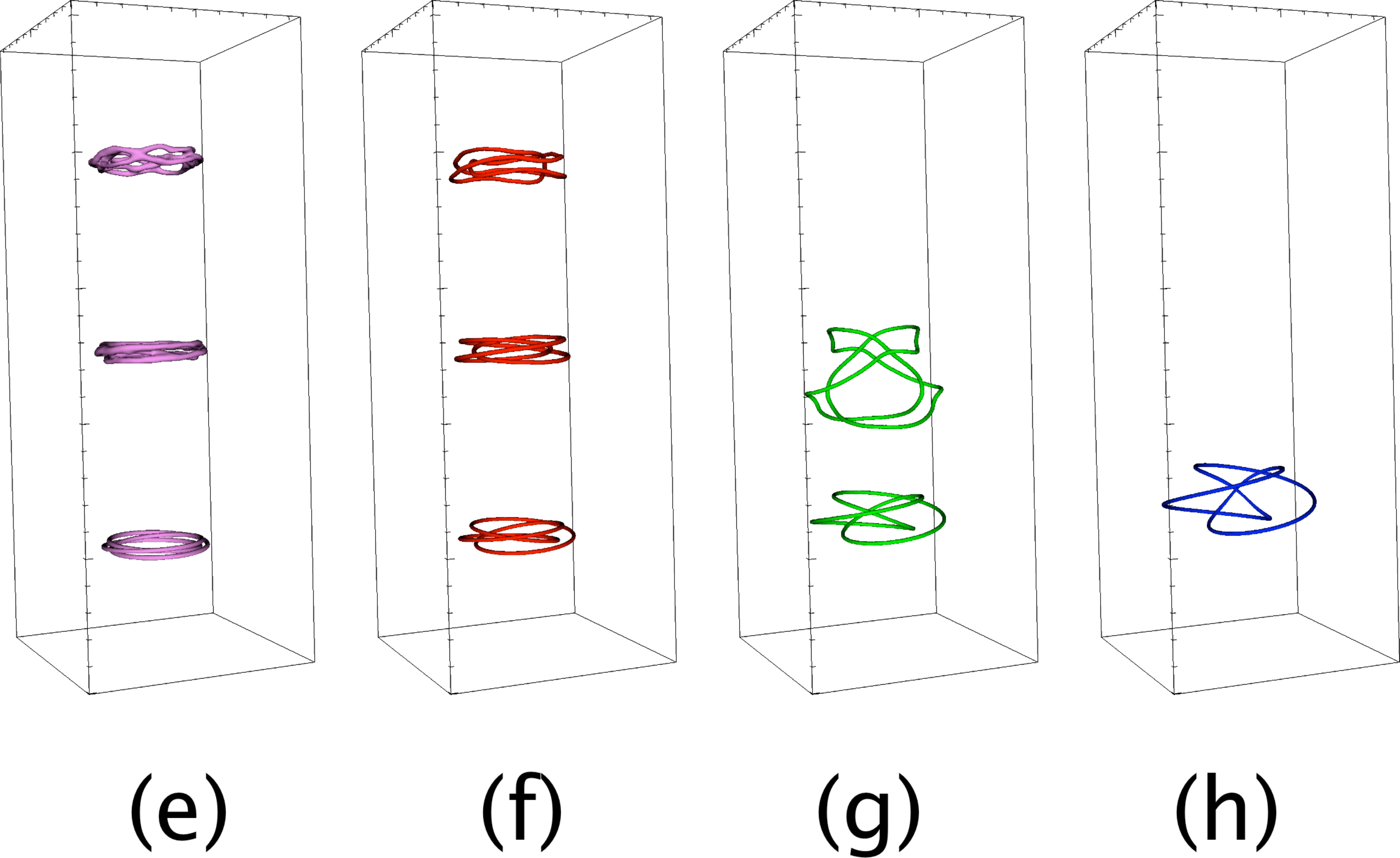}
\caption{(Colours online).
Iso-surfaces of the density field 
at the threshold level $ \rho_{th}=0.2 $ for  $ \mathcal{T}_{3, 2} $ 
knots of various knot ratios $ R_1/R_0 $ (see Table 
\ref{tab:param}).
Snapshots at times to $ t=0, 400,
800 $. Unstable knots are not shown.
\label{fig:evolutionT32}}
\end{figure}

Fig. \ref{fig:evolutionT23} and \ref{fig:evolutionT32}  
show the iso-surfaces of the density field corresponding to the 
threshold value $ \rho_{th}=0.2 $ at the initial conditions and at
successive times for the  $ \mathcal{T}_{2, 3} $ and 
$ \mathcal{T}_{3, 2} $ knots respectively (unstable knots will
be discussed in the next section).
As expected, vortex knots move along the z-direction (the
axis of symmetry of the torus), but also twist
around it.
Qualitatively, vortex knots with small knot ratio $ R_1/R_0 $ 
are fast and stable, as they propagate along the z-direction without
breaking.
During the evolution, Kelvin waves \cite{donnelly1991} 
appear; such waves are visible at the last stages
of cases (a), (b), (e), and (f).

\begin{figure}
\includegraphics[scale=1.3]{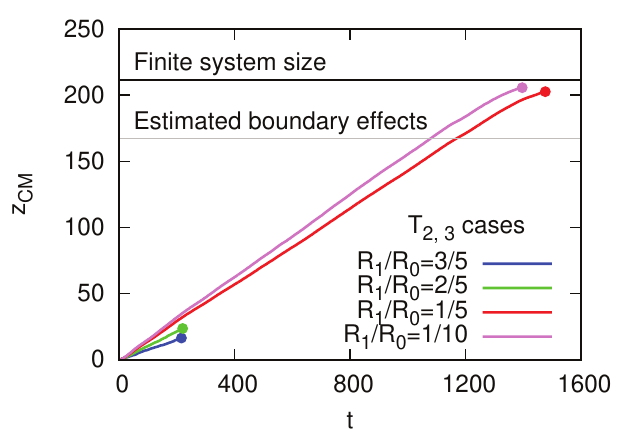}
\caption{
(Colours online).
Position along the z axis (in units of the healing length)
of the centre of mass of
$ \mathcal{T}_{2, 3} $ knots of various knot ratios 
$ R_1/R_0 $ as a function of time. 
The filled circles denote the position where vortex knots
break up. The horizontal lines denote respectively the distance
where boundary effects become non-negligible and the
finite system size along z.
\label{fig:knotT23PositionsZ}}
\end{figure}

\begin{figure}
\includegraphics[scale=1.3]{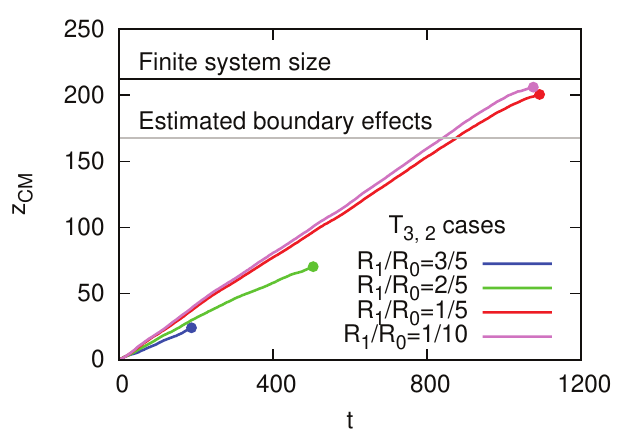}
\caption{(Colours online).
As in Fig.~\ref{fig:knotT23PositionsZ} but for
$ \mathcal{T}_{3, 2} $.
\label{fig:knotT32PositionsZ}}
\end{figure}

In order to quantify the evolution of vortex knots and compare one
knot with others, we define
the knot center of
mass $ \mathbf{r}_{CM}=(x_{CM}, y_{CM}, z_{CM}) $  as
\begin{equation}
\mathbf{r}_{CM}=\frac{\int_V \mathbf{r} \, H\left(\rho_{th}-|\psi|^2\right) \, dV}{\int_V H\left(\rho_{th}-|\psi|^2\right) dV},
\end{equation}
where $ H(...) $ is the Heaviside step function.
Fig. \ref{fig:knotT23PositionsZ} and  \ref{fig:knotT32PositionsZ}
show the $ z $ component $ z_{CM} $ of the knot center 
of mass (shifted with respect to the initial position) 
for the $ \mathcal{T}_{2, 3} $ and $ \mathcal{T}_{3, 2} $
cases respectively.
In both cases, knots with smaller knot 
ratio $ R_1/R_0 $ move faster, are more stable 
and propagate for longer distances before breaking up
(a filled symbol at the end of each curve marks the break-up point).

\begin{figure}
\includegraphics[scale=1.3]{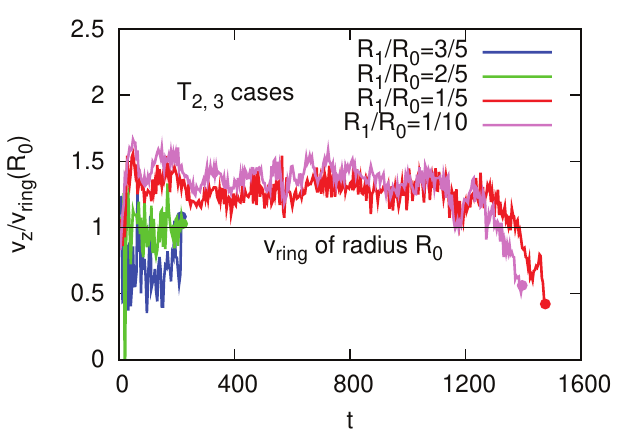}
\caption{(Colours online).
Velocity component $ v_z $ of  $ \mathcal{T}_{2, 3} $ versus time of
knots with various knot ratios $ R_1/R_0 $ before 
destroying (filled points).
Velocities are expressed in units of vortex ring velocity (\ref{eq:ring})
with quantum number $ n=1 $ and radius $ R=R_0 $.
\label{fig:knotT23VelocitiesZ}}
\end{figure}

\begin{figure}
\includegraphics[scale=1.3]{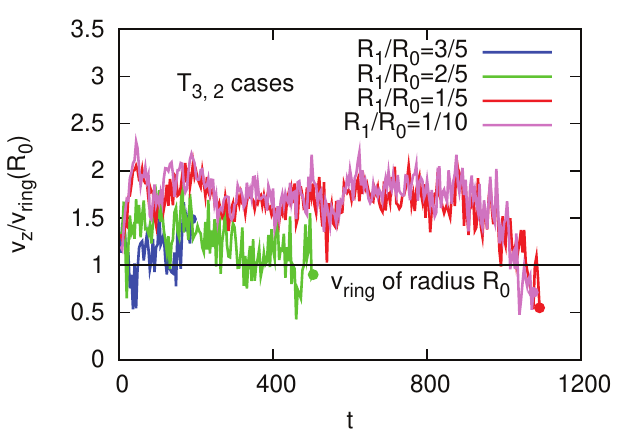}
\caption{(Colours online).
Velocity component $ v_z $ of  $ \mathcal{T}_{3, 2} $ 
knots having different knot ratio $ R_1/R_0 $ before 
destroying (filled points).
Velocities are expressed in units of ring velocity (\ref{eq:ring})
having quantum number $ n=1 $ and radius $ R=R_0 $.
\label{fig:knotT32VelocitiesZ}}
\end{figure}

The z component of the velocity of a vortex knot is estimated
by evaluating
$ v_z(t) = \left[z_{CM}(t+\tau) 
-z_{CM}(t) \right] / \tau $ (where $ \tau=4 $ for numerical
convenience).
Fig. \ref{fig:knotT23VelocitiesZ} and Fig. 
\ref{fig:knotT32VelocitiesZ} show $ v_z(t) $ 
measured in units of the vortex ring velocity \cite{roberts1971}
\begin{equation}
v_{\mbox{ring}}(R) = \frac{n\, \kappa}{4\pi R} \left[ \ln \left(\frac{8R}{\xi}\right)-0.615\right]
\label{eq:ring}
\end{equation}
having quantum number $ n=1 $ and radius $ R=R_0 $
(note that in our non-dimensional system the quantum of
circulation is $ \kappa = 2\pi $).
It s apparent that vortex knots move with approximately constant
$ z $ velocity before either breaking up or reaching the 
boundary of the computational domain, where the interaction with the 
image slows them down.

\begin{figure}
\includegraphics[scale=1.3]{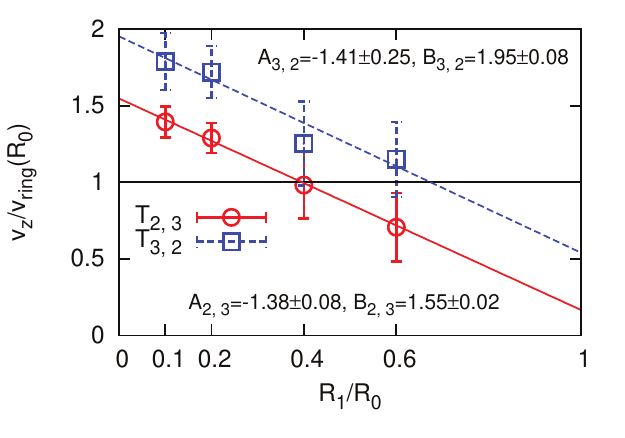}
\caption{
(Colours online).
Averaged velocity components $ v_z $ of  $ \mathcal{T}_{2, 3} $
and $ \mathcal{T}_{3, 2} $ vortex knots with various
knot ratios $ R_1/R_0 $.
Velocities are expressed in units of vortex ring velocity (\ref{eq:ring})
with quantum number $ n=1 $ and radius $ R=R_0 $; error-bars 
correspond to one standard deviation.
\label{fig:knotVelocityZScalings}}
\end{figure}

It is instructive to analyze the mean and the standard deviation of
the vortex knots' velocities measured in the constant-velocity regimes.
The results, expressed in units of $ v_{\mbox{ring}}(R_0) $, are 
shown in Fig. \ref{fig:knotVelocityZScalings}.
Three conclusions can be drawn from this figure:
\begin{enumerate}

\item
$\mathcal{T}_{2, 3} $ knots are slower than 
$ \mathcal{T}_{3, 2} $ knots with the same knot ratio.
This is physically expected as the velocity field of torus knots at
large distance is similar to the velocity field of vortex rings
with multiple circulation:
$ \mathcal{T}_{2, 3} $ corresponds to circulation
of $ 2\kappa $ and $ \mathcal{T}_{3, 2} $ to $ 3\kappa $.
According to Equation (\ref{eq:ring}), the velocity is directly
proportional to the circulation, and so $ \mathcal{T}_{2, 3} $ knots
should be slower than $ \mathcal{T}_{3, 2} $ ones.
However, this simple consideration does not apply well to
knots because we would have expected, for the small knot
ratio tested ($ R_1/R_0=1/10 $) a scaled velocity of 
$ v_z/v_{\mbox{ring}}(R_0)\simeq2 $ and 
$ v_z/v_{\mbox{ring}}(R_0)\simeq3 $ for $ \mathcal{T}_{2, 3} $ 
and $ \mathcal{T}_{3, 2} $ respectively, and this is not the case.

\item
The z-velocity component scales with the knot ratio and can be parametrised as
\begin{equation}
v_z \left(\frac{R_1}{R_0}\right)=A_{n, m} \frac{R_1}{R_0} + B_{n, m}
\end{equation}
where $ A_{n, m} $ and $ B_{n, m} $ are coefficients which refer to the
generic torus knot $ \mathcal{T}_{n, m}$.
Values of $ A_{n, m} $ and $ B_{n, m} $ for the knots
$ \mathcal{T}_{2, 3} $ and $ \mathcal{T}_{3, 2} $ are reported in
Fig. \ref{fig:knotVelocityZScalings}.
It is interesting to observe that $ A_{2, 3}\simeq A_{3, 2} $.

\item
The z-velocity component of unstable knots (i.e. knots
that decay before reaching the computational boundaries)
is less or similar to $ v_{\mbox{ring}}(R_0) $.
On the contrary, knots that remain stable within the computational domain 
are characterized by $ v_z > v_{\mbox{ring}}(R_0) $.
\end{enumerate}

%\clearpage

\section{The breaking of a knot \label{sec:BC}}
In our simulations we have observed that some vortex knots break up
into topologically simpler objects.

\begin{figure}
\includegraphics[scale=0.12]{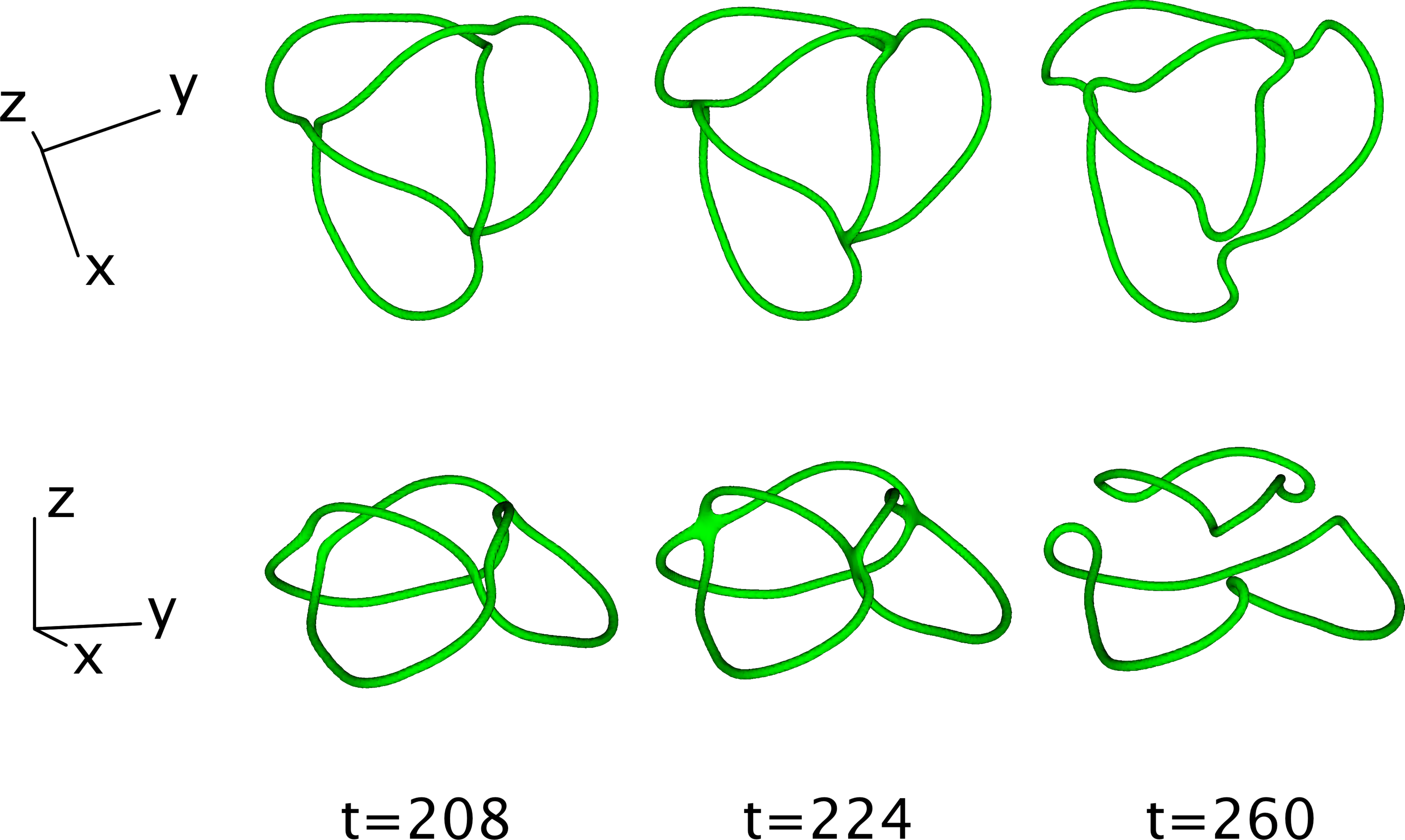}
\caption{(Colours online).
Three successive snapshots showing how the $ \mathcal{T}_{2, 3} $
vortex knot with knot ratio $ R_1/R_0=2/5 $ breaks up into two vortex rings.
Here we plot two perspectives (up and side to the vortex propagation) of
the iso-surfaces of the density field corresponding to the threshold level
$ \rho_{th}=0.2 $.
\label{fig:destructionT23}}
\end{figure}

We first analyze the unstable $ \mathcal{T}_{2, 3} $ knots: these are knots
corresponding to knot ratios $ R_1/R_0=2/5, 3/5 $.
In Fig. \ref{fig:destructionT23} we show three snapshots of the 
decay of the $ \mathcal{T}_{2, 3} $ knot with ratio $ R_1/R_0=2/5 $.
It is apparent that the knot breaks into two vortex rings
via three simultaneous self-reconnection events (see in particular the snapshot
corresponding to $ t=224 $).
The decay of the vortex knot $ \mathcal{T}_{2, 3} $ 
with ratio $ R_1/R_0=3/5 $, 
not shown here, is similar.

\begin{figure*}
\includegraphics[scale=0.068]{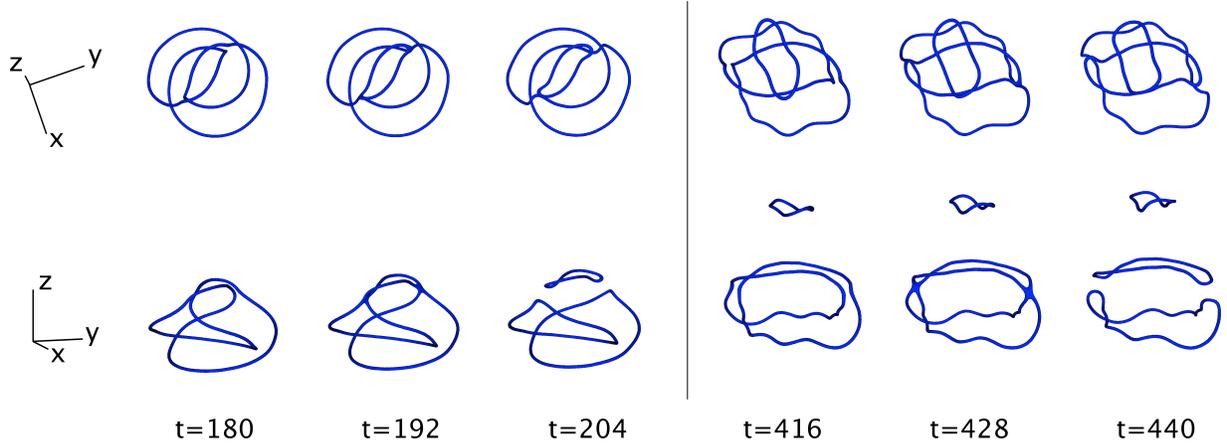}
\caption{(Colours online).
Two sequences of three successive snapshots showing how the 
$ \mathcal{T}_{3, 2} $ vortex knot with knot ratio $ R_1/R_0=3/5 $ breaks 
up into 
three vortex rings.
Here we plot two perspectives (up and side to the vortex propagation) of
the iso-surfaces of the density field corresponding to the threshold 
level $ \rho_{th}=0.2 $.
\label{fig:destructionT32_3-5}}
\end{figure*}

On the contrary, $ \mathcal{T}_{3, 2} $ vortex knots break in a different
manner.
As shown in Fig. \ref{fig:destructionT32_3-5}, the vortex knot $ \mathcal{T}_{3, 2} $
with knot ratio $ R_1/R_0=3/5 $ first decays in one vortex ring and two linked
vortex rings via two simultaneous self-reconnection events (snapshot at
time $ t=192 $).
Subsequently, the small free ring escapes from the other rings, which
undergo two simultaneous reconnection events
that create two unlinked vortex rings (snapshot at time $ t=428 $).
The last step is remarkable:
there is no apparent reason why two linked vortex knots should in principle
unlink into two vortex rings (by making two simultaneous reconnection
events) without forming a single ring (by one reconnection event).

\begin{figure}
\includegraphics[scale=0.136]{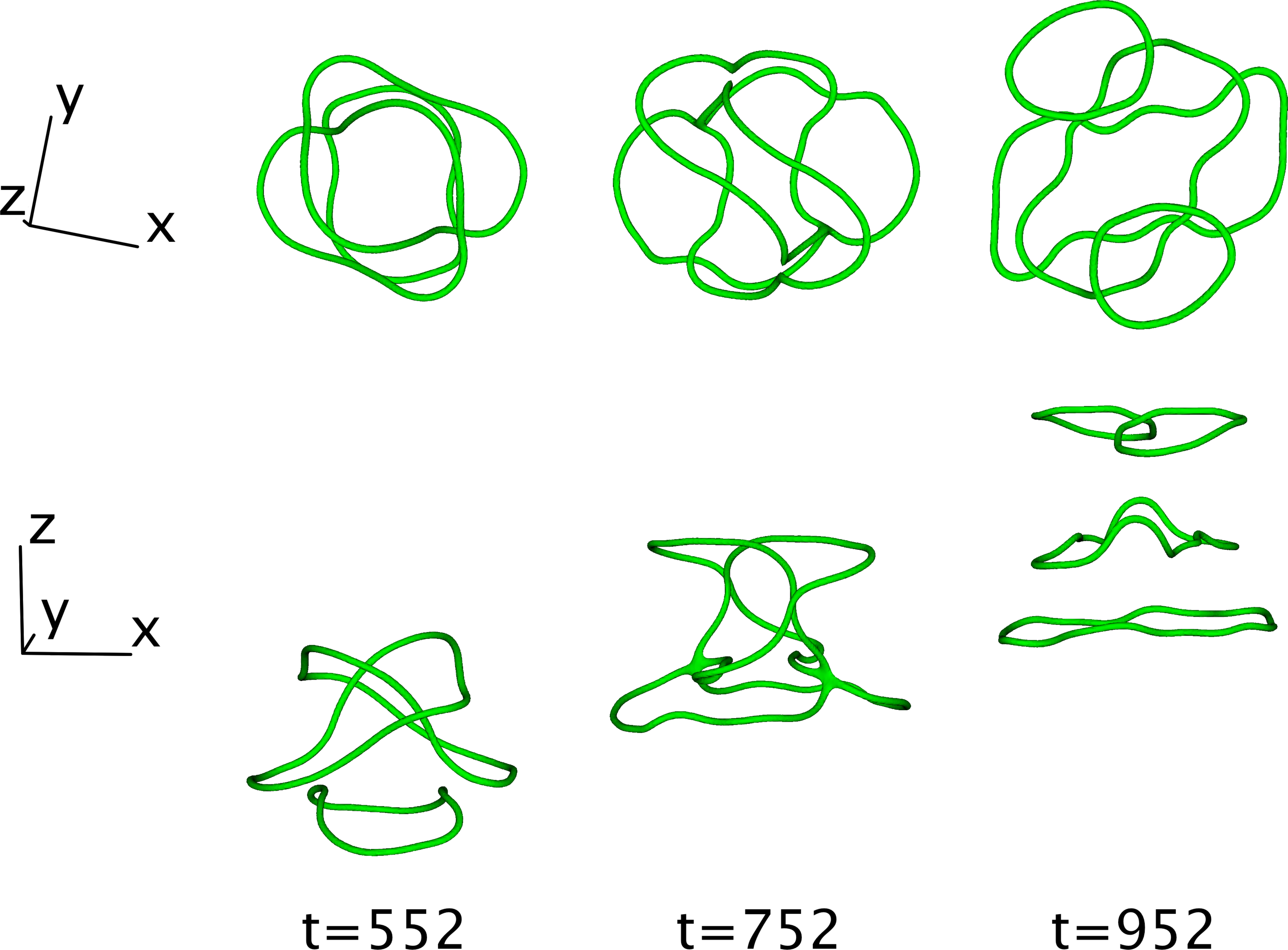}
\caption{(Colours online).
Three successive snapshots showing how the $ \mathcal{T}_{3, 2} $
knot with knot ratio $ R_1/R_0=2/5 $ breaks into three vortex rings.
Here we plot two perspectives (up and side to the vortex propagation) of
the iso-surfaces of the density field corresponding to the threshold $ \rho_{th}=0.2 $.
\label{fig:destructionT32_2-5}}
\end{figure}

The $ \mathcal{T}_{3, 2} $ vortex knot with ratio $ R_1/R_0=2/5 $ qualitatively 
decays in the same way, producing a set of three unlinked
vortex rings, but the steps are quite different.
In the first step, a free vortex ring and two linked vortex rings
are again produced.
However, the free ring, which is initially located behind the two 
linked vortex rings, is smaller and faster than the other rings.
As a consequence, it reconnects with the two linked vortex rings,
as shown in Fig. \ref{fig:destructionT32_2-5} (snapshot $ t=752 $).
At this point the reformed knot breaks up, undergoing the same sequence 
previously described for the $ \mathcal{T}_{3, 2} $ with ratio 
$ R_0/R_1=3/5 $ case, and the outcome is a set of three vortex rings 
(snapshot at time $ t=952 $).
Note that, in the last snapshot, the first knot 
(in the sense of the position) 
has split into two smaller vortex rings via
a self-reconnection 
event which is probably consequence of its Kelvin waves oscillations.

\section{Conclusions \label{sec:CONCL}}
We have numerically analyzed the existence and stability of
vortex knots in the GPE model of a condensate.
We have proposed a novel numerical technique for creating 
{\it ab initio} vortex knots in the wave-function of the condensate.
In particular, we have focussed our numerical computations on the two simplest
(in the topological sense) vortex knots, $ \mathcal{T}_{2, 3} $ 
and the $ \mathcal{T}_{3, 2} $.
We have analyzed the evolution and the stability of such knots 
with respect to the knot ratio $ R_1/R_0 $.
We have found that a knot can be unstable, i.e. it breaks up into 
simple rings during the propagation, or stable within our computational 
domain. 
Our numerical experiments clearly show that a small knot ratio 
($ R_1/R_0=1/10, 1/5 $) increases the stability, whereas a large
knot ratio 
($ R_1/R_0=2/5, 3/5 $) decreases it, in agremeent with \cite{ricca:1999}.

We have found that vortex knots propagate essentially as vortex rings.
We have measured the vortex knot velocities along the torus 
symmetry axis and shown that the
velocity depends linearly on the knot ratio for both
$ \mathcal{T}_{2, 3} $ and $ \mathcal{T}_{3, 2} $.

Finally, we have studied the details of the break up of vortex knots.
Although we do not have a theoretical explanation for the break up,
we have observed evidences of
generic breaking behavior:
$ \mathcal{T}_{2, 3} $ 
knots always break into two vortex rings via a three 
simultaneous self-reconnection event, whereas.
$ \mathcal{T}_{2, 3} $ knots first decay into 
three vortex rings via two simultaneous self-reconnections
which create a free ring and two linked 
rings, then undergo two simultaneous reconnections 
which split the resulting link.

We believe that our work opens up new interesting problems
in the field of fluid topology applied to superfluids and 
 Bose-Einstein condensates. The natural developments of our
study will be a
theoretical investigation of the stability of vortex knots, and
an experimental study of the creation of a knotted initial
condition in an atomic condensate.

\begin{acknowledgments}
The authors acknowledge G. Boffetta, F. De Lillo, A.L. Fetter and A.J. Youd
for comments and suggestions. CFB is grateful to the Leverhulme Trust for
financial support.
The authors also thank developers at
 {\it LLNS VisIt}, {\it POV-Ray} and {\it Gnuplot} for the
free software used for  the visualization of numerical results.
\end{acknowledgments}

\bibliography{references}

\end{document}